\documentclass[12pt]{article}

\usepackage[utf8]{inputenc}
\usepackage[T1]{fontenc}
\usepackage{amsmath,amssymb}
\usepackage{graphicx}
\usepackage{booktabs}
\usepackage{array}
\usepackage{longtable}
\usepackage[hidelinks]{hyperref}
\usepackage[margin=1in]{geometry}
\usepackage{caption}
\usepackage{subcaption}
\usepackage{float}
\usepackage{multirow}
\usepackage{tabularx}
\usepackage{cite}
\usepackage{adjustbox}

\graphicspath{{figures/}}

\title{\textbf{A Novel Approach for Testing Water Safety Using Deep Learning Inference of Microscopic Images of Unincubated Water Samples}}
\author{Sanjay Srinivasan\\
\texttt{sanjasrin@gmail.com}\\
Eastlake High School}
\date{}

\begin{document}

\maketitle

\newpage

\begin{abstract}
Fecal-contaminated water causes diseases and even death. Current microbial water safety tests require pathogen incubation, taking 24-72 hours and costing \$20-\$50 per test. This paper presents a solution (DeepScope) exceeding UNICEF's ideal Target Product Profile requirements for presence/absence testing, with an estimated per-test cost of \$0.44. By eliminating the need for pathogen incubation, DeepScope reduces testing time by over 98\%.

In DeepScope, a dataset of microscope images of bacteria and water samples was assembled. An innovative augmentation technique, generating up to 21 trillion images from a single microscope image, was developed. Four convolutional neural network models were developed using transfer learning and regularization techniques, then evaluated on a field-test dataset comprising 100,000 microscope images of unseen, real-world water samples collected from fourteen different water sources across Sammamish, WA. Precision-recall analysis showed the DeepScope model achieves 93\% accuracy, with precision of 90\% and recall exceeding 94\%. The DeepScope model was deployed on a web server, and mobile applications for Android and iOS were developed, enabling Internet-based or smartphone-based water safety testing, with results obtained in seconds.
\end{abstract}

\newpage

\tableofcontents

\newpage

\section*{Keywords}
Fecal contamination, coliform bacteria, presence/absence test, microscopy, convolutional neural network, UNICEF ideal Target Product Profile

\section*{Abbreviations}
\begin{tabular}{|l|l|}
\hline
WHO & World Health Organization \\
\hline
CDC & Center for Disease Control \\
\hline
E. Coli & \textit{Escherichia Coli} \\
\hline
US EPA & US Environmental Protection Agency \\
\hline
CNN & Convolutional Neural Network \\
\hline
UNICEF TPP & United Nations International Children's Emergency Fund Target Product Profile \\
\hline
ASTM & American Society for Testing and Materials \\
\hline
DIBaS & Digital Images of Bacteria Species \\
\hline
\end{tabular}

\section{Introduction and Related Work}

As of 2022, two billion people in the world did not have access to safe water \cite{un_water}. One of the leading sources of unsafe water is water contaminated with feces. WHO estimates that at least 1.7 billion people were drinking water contaminated with feces in 2022. Fecal contamination can introduce harmful pathogens such as viruses or bacteria into the water supply. These pathogens can cause a variety of waterborne diseases such as diarrhea, cholera, typhoid, dysentery, polio, and hepatitis, which cause over 505,000 deaths each year \cite{who_drinking}. In the USA, exposure to bacteria in drinking water causes 40\% of hospitalizations and 50\% of deaths related to waterborne illnesses. This costs the USA \$1.39 billion each year \cite{cdc_surveillance}.

Globally, 1.7 billion people consume water contaminated with feces each year \cite{who_drinking}. Fecal contamination in water can occur as a result of extreme weather and climate events, which may lead to farm waste spilling into water supplies, or as a result of inadequate access to proper sanitation facilities. Fecal contamination affects both high-income as well as low- and middle-income countries, making it a global issue \cite{epa_fecal, who_sanitation, epa_ecoli}. Boiling water before consumption remains the most effective safeguard against such contamination. However, this does not protect against skin contact through swimming, showering, wading in the water, etc. It is therefore important to detect fecal contamination to protect human health. E. coli is the most feasible and cost-effective indicator to determine water quality for fecal contamination and is used as an indicator organism by international drinking water guidelines and national regulations \cite{pichel2023}. The most commonly used approaches for detecting bacterial contamination include presence/absence tests, and laboratory-based testing (e.g., membrane filtration) \cite{unicef_tpp}.

Presence/absence testing is a widely used method for detecting microbial contamination in water in resource constrained environments. It is intended to be a first line of defense and does not require sophisticated skills of the user.

Presence/absence tests indicate whether coliform bacteria are present in a given water sample. Conventionally, this test involves incubating the pathogens in a water sample for 24-72 hours, allowing them to grow and produce an enzyme that when interacted with, produces a color change. These tests are inexpensive, with a commercial cost of $\sim$20 dollars. On average, current presence/absence tests (such as Colilert-18, Coliscan, Readycult) have a failure rate of 0-23.3\%. Most tests fall within UNICEF's acceptable range of a less than 15\% false positive and false negative rate \cite{unicef_tpp, olstadt2007}.

Laboratory based water tests provide high accuracy and quantitative microbial counts. These tests involve passing a water sample through a membrane filter, which retains the water sample's microbial contaminants. The filter is then placed on a growth medium and incubated for 18-24 hours at high temperatures (35$^\circ$- 44.5$^\circ$ C) to allow bacterial colonies to grow and develop. These tests have a commercial cost of $\sim$50 dollars. Membrane filtration provides direct microbial counts and achieves a false positive and false negative rate of approximately 4.3\%, which falls within UNICEF's ideal range. This method of testing requires a laboratory as well as trained scientists, which can limit its use in field testing applications \cite{astm_ecoli, bain2012, epa_method1604}.

WaterScope \cite{dabrowska2024} is an alternative to traditional laboratory testing methods. This product involves incubating the bacteria for 8-21 hours through a growth medium. A microscopic image is taken and afterward analyzed through a machine learning algorithm. WaterScope provides quantitative microbial counts after 21 hours, and presence/absence results within 8 hours. WaterScope does not require a laboratory, allowing for use in resource-constrained areas. However, on a field test of 130 samples, WaterScope could detect bacterial contamination with an accuracy of only 67\% within 8 hours. This falls well below UNICEF's TPP \cite{unicef_tpp} for water safety tests.

According to UNICEF's TPP, the ideal presence/absence test would provide results in under 30 minutes to help ensure quick action to be taken, be able to be used in field settings, and have a cost of under six dollars per test. On a testing set of at least 30 samples, this test should have a false positive and false negative rate less than 10\% on a minimum sample at 90\% confidence level. The ideal presence/absence would not require the use of reagent mixing or incubation \cite{unicef_tpp}. Currently, to the best of the author's knowledge, there is no solution that meets these criteria (the work described in this paper is a step in this direction).

Whether using a presence/absence test or a laboratory test, detecting water contamination using current methods with acceptable accuracy is a time-consuming and potentially expensive exercise. Thus, a cost-effective, fast, and accurate solution to test water safety is needed. In this paper, one potential solution to this problem and the experimental validation of this solution are discussed. Specifically, microscopy and deep learning are combined to create models that predict whether a microscopic image of an unincubated water sample represents a safe or unsafe water source. This paper also discusses how these deep learning models can be made widely available using the Internet and smartphone applications.

Smartphones are a particularly powerful tool that are available to billions of users across the world even in poor countries that are otherwise constrained by infrastructure. Smartphones can be used to take images of water samples under a microscope, and the image tested using the deep learning model, either on the smartphone itself or over the Internet.

\section{Engineering Goals}

\noindent\textbf{Goal 1:} Develop high performing deep learning models to test water safety that adhere to accepted guidelines. These models will utilize microscopic images of unincubated samples, ensuring quick and accurate contamination detection.

\noindent\textbf{Goal 2:} Develop a webserver and smartphone application to host and distribute deep learning models via the Internet. This will allow users to assess water sample safety remotely online through smartphones.

\noindent\textbf{Goal 3:} Develop a smartphone application designed to distribute deep learning models directly on the device, enabling water safety detection in Internet-constrained areas.

\section{Methodology: Microscopy of Water Samples}

\begin{figure}[H]
\centering
\includegraphics[width=0.5\textwidth]{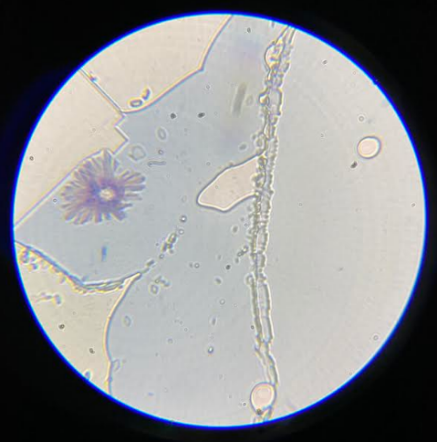}
\caption{Image of Rainwater Puddle Underneath Compound Microscope at 1000x Magnification}
\label{fig:rainwater}
\end{figure}

There is a wide range of relatively inexpensive microscopes available on the market today. In this project, three different inexpensive microscopes were tested for suitability for examining bacteria: the Carson MicroFlip with an advertised magnification of 250x, a USB microscope with advertised magnification of 1600x, and an AmScope M150 with a range of optical magnifications up to 1000x. It was determined that the Carson MicroFlip and the USB microscope were unsuitable for viewing bacteria. The AmScope 150 was determined to be suitable for viewing bacteria, preferably at 1000x magnification. In these initial experiments water from a rainwater puddle (which tested positive for coliform bacteria in a color-change coliform test) was used. To improve potential viewability, some of the water samples were also stained with methylene blue dye. An example image of a microscopic organism from a rainwater puddle at 1000x magnification is shown in Figure \ref{fig:rainwater}.

To standardize the microscope examination, a protocol was developed. For each water sample, a drop of water was placed on two separate slides using a pipette. Methylene blue dye was used to stain one of the drops of water. A cover slide was then placed on each slide, and the water was allowed to dry. At room temperature, the water drops typically dried in under 5 minutes. The slide was then examined under the microscope. Photos of the microscopic view were taken by placing a cell phone camera above the eyepiece of the microscope. The protocol was developed in a manner that is easy to follow by eventual users of the solution being developed in this project.

In the testing protocol adopted for this project, the water samples were immediately discarded after microscopic examination (except for the samples under coliform color change testing) to reduce exposure to potentially contaminated water.

\section{Methodology: Deep Learning Dataset Construction}

\subsection{Foundational Dataset}

Training deep learning models uses a substantial amount of labeled data for each class to achieve accurate predictions. To begin creating a foundational dataset, the publicly available DIBaS dataset was used.\footnote{The image dataset URL provided in Deep Learning Approach to Bacterial Colony Classification, 2017 no longer works. However, a copy of this dataset is available on Github at \url{https://github.com/gallardorafael/DIBaS-Dataset}. The DIBaS images from Github were downloaded.} DIBaS is a collection of microscopic images of bacteria taken from an Olympus CX31 Upright Biological Compound Microscope. DIBaS contains 660 images spanning 33 different bacterial species. Images of three commonly occurring and dangerous types of bacteria in water contaminated with feces, \textit{Enterococcus Faccieum}, \textit{Enterococcus Faecalis}, and \textit{E. Coli}, were added to the foundational dataset. Additionally, various strains of \textit{Lactobacillus}, a genus of safe bacteria for humans, were included. These strains include \textit{Lactobacillus Casei}, \textit{Lactobacillus Crispatus}, \textit{Lactobacillus Delbrueckii}, \textit{Lactobacillus Gasseri}, \textit{Lactobacillus Jehnsenii}, \textit{Lactobacillus Johnsonii}, \textit{Lactobacillus Paracasei}, \textit{Lactobacillus Plantarum}, \textit{Lactobacillus reuteri}, \textit{Lactobacillus Rhamnosus}, and \textit{Lactobacillus Salivarius} \cite{zielinski2017}.

\begin{figure}[H]
\centering
\includegraphics[width=0.4\textwidth]{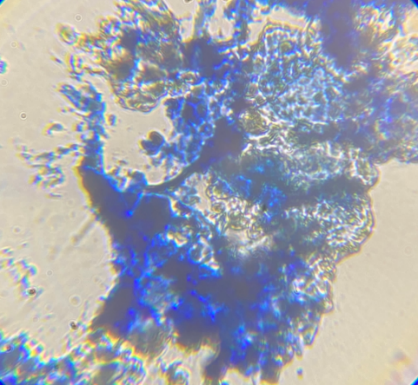}
\caption{Image of Shower Drain Water Underneath Compound Microscope Stained with Methylene Blue Dye at 1000x Magnification.}
\label{fig:shower_drain}
\end{figure}

While the DIBaS dataset is comprehensive and includes detailed images obtained from a high-resolution microscope, it also makes the data set inaccessible for practical applications where image samples maybe taken from less accurate microscopes. Therefore, there was a need to include image samples that were more error-prone and taken with lower resolution microscopes.

Additional microscopic images of water samples were added to the foundational dataset.\footnote{\url{https://github.com/sanjasrin/DeepScope}} These images were created by taking photos of the sample under the microscope using a cellphone camera. Three distinct types of water samples were included to expand the dataset: (1) clean, bottled drinking water purchased from a local grocery store, (2) water from a shower drain, and (3) water from a rainwater puddle. Both the clean bottled water and the shower drain water tested negative for coliform bacteria in a coliform color change test, whereas the rainwater puddle tested positive for coliform bacteria. Despite the absence of coliform bacteria in the shower drain water, a significant amount of particulate matter was observed, leading to the categorization of the water sample as unsafe. The foundational dataset comprised 297 microscopic images.

\subsection{Augmentations to Dataset}

To enhance the robustness and size of the dataset, augmentation techniques were applied to the images. With smaller datasets, there is a high likelihood of overfitting. Overfitting occurs when a deep learning model performs extremely well on the training data, but poorly on validation and testing. The model will recognize specific details and patterns in a small dataset and memorize the training data rather than identifying generalized trends and patterns. Applying augmentations to the dataset is one way to fix overfitting. Augmentations allow for the deep learning model to extract more information from the original dataset, making the dataset substantially larger through techniques like data warping or oversampling, leading to better and more accurate deep learning models \cite{shorten2019}.

\begin{figure}[H]
\centering
\includegraphics[width=0.7\textwidth]{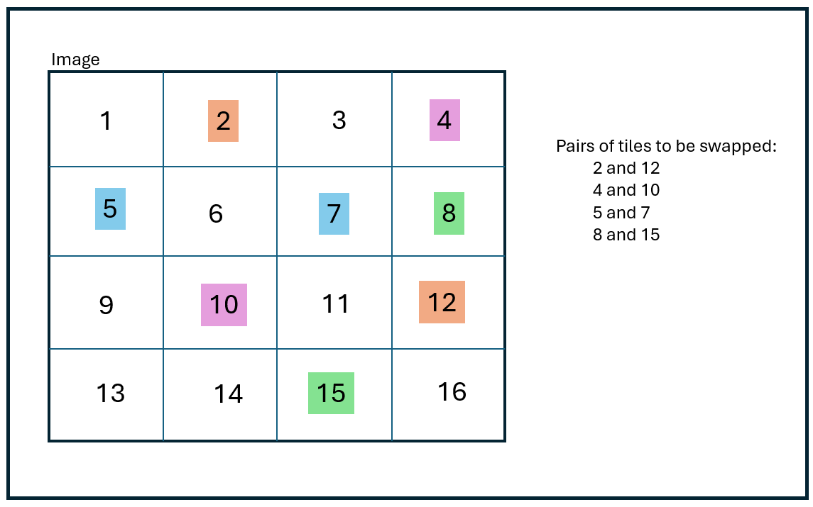}
\caption{Illustration of image augmentation. The image is divided into 16 tiles, and random pairs of tiles are selected and their contents swapped.}
\label{fig:augmentation}
\end{figure}

A novel method of data augmentation was developed to address the specific challenges within this problem space, as illustrated in Figure \ref{fig:augmentation}. Through analysis of microscopic images of bacteria and water samples, it was observed that relocating portions of an image, particularly regions containing bacteria, to different locations within the image did not alter its representation as a safe or unsafe water sample. Building on this observation, data tiles within the image were systematically identified and rearranged to produce augmented images that retained the essential information of the original sample. To perform these augmentations, input images were first converted into square formats, enabling them to be partitioned into equal-sized square tiles for further processing.

A 4x4 tile pattern was selected as it preserves the ability to recognize bacterial patterns, unlike smaller tile configurations (e.g., 32x32), where certain details and patterns might be lost. The 4x4 tile pattern allows for the generation of a vast number of images (16! or 20,922,789,888,000 combinations). Pairs of tiles were randomly selected and swapped with each other, creating new arrangements of tiles while preserving the original patterns of the water sample in each tile. The number of tile pairs to be swapped was also randomly determined (ranging between four and twelve), allowing for variety in the augmented images without completely altering the new image. A total of 30,000 new images were created---approximately 15,000 for each class, representing a 100-times increase in the size of the original dataset. This substantial increase in data diversity potentially aids in reducing overfitting.

\subsection{Field Testing Dataset}

In evaluating water safety testing methods in field environments, multiple studies have been conducted. Generally, water has been sampled from 3-16 water sources. These sample sizes reflect the practical range used in field-testing of water quality \cite{olstadt2007, brown2020, nam2014, pisciotta2002}. For evaluating the deep learning models developed in this paper, a field-test set of 160 microscope images from 14 different water sources was constructed.

\begin{figure}[H]
\centering
\includegraphics[width=0.6\textwidth]{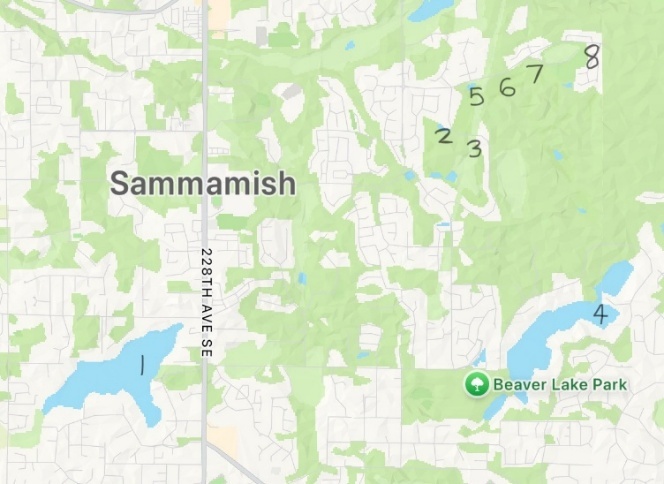}
\caption{Locations of water bodies in Sammamish where water samples were collected for field testing.}
\label{fig:map}
\end{figure}

The objective was to test the deep learning models developed in this paper against water samples that had not been encountered during the model's development. The performance of the model against these water sample images provides data on the accuracy and robustness of the model when applied to unseen data, representing the conditions the model would face when deployed for real-world use. Of the 14 different water sources, 5 were clean water sources from different brands of drinking water. One sample was from tap-water. The remaining 8 water sources were various water bodies in Sammamish, Washington. The water sources are shown in Figure \ref{fig:map}.

\begin{figure}[H]
\centering
\begin{subfigure}[b]{0.45\textwidth}
\centering
\includegraphics[width=\textwidth]{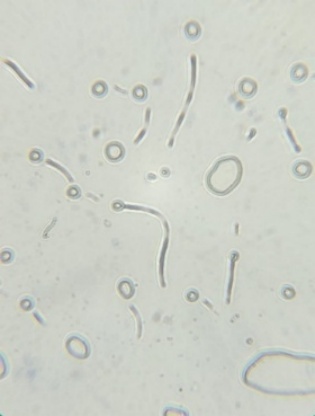}
\caption{Source 5}
\end{subfigure}
\hfill
\begin{subfigure}[b]{0.45\textwidth}
\centering
\includegraphics[width=\textwidth]{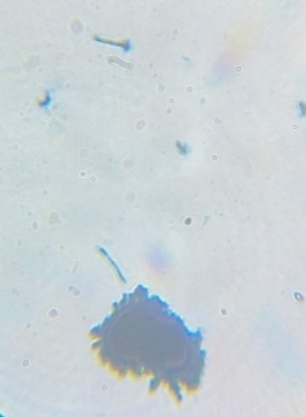}
\caption{Source 6}
\end{subfigure}
\caption{Microscope images at 1000x magnification of water sample from Sources 5 (left) and 6 (right). Both sources were found to be unsafe in color change coliform testing.}
\label{fig:sources56}
\end{figure}

After collecting these water samples\footnote{Water samples were collected from various water bodies. Chemical reagent testing was conducted on-site at the water body by collecting samples into a container and inserting test strips directly into the water. The samples were discarded immediately after capturing microscope images, except for the coliform color change testing samples, which were disposed of upon completion of the testing.}, images of the samples under a microscope were taken at 1000x optical magnification. Figure \ref{fig:sources56} shows sample microscope images obtained from the water samples. Microscopic examinations of each water sample were conducted, both with and without methylene blue dye. 10-12 microscopic images were taken of samples from each water source. Thus, the same water sample generated different looking images (same as with the foundational data set). All 160 microscopic images created have been uploaded to Github.\footnote{\url{https://github.com/sanjasrin/DeepScope}} A coliform color change test was performed on each water sample, as well as a chemical reagent test to test the presence of unsafe chemicals. The chemical reagent test consists of a strip coated with reagents at different locations that change various colors when dipped in water to indicate the presence of different types of chemicals such as nitrites, fluorides, mercury, etc. All the water samples tested negative for the presence of unsafe chemicals, but 6 of the water sources tested positive for coliform bacteria. These results are summarized in Table \ref{tab:water_sources}.

\begin{table}[H]
\centering
\caption{Details of water samples from field testing.}
\label{tab:water_sources}
\footnotesize
\adjustbox{max width=\textwidth}{
\begin{tabular}{|c|l|c|c|c|c|}
\hline
\textbf{Source} & \textbf{Name} & \textbf{Coliform Test} & \textbf{Chemical Test} & \textbf{Safe/Unsafe} & \textbf{Images} \\
\hline
1 & Pine Lake & Negative & Negative & Safe & 12 \\
\hline
2 & Pond 1 Windsor Greens & Positive & Negative & Unsafe & 12 \\
\hline
3 & Windsor Greens Fountain & Positive & Negative & Unsafe & 12 \\
\hline
4 & Beaver Lake & Positive & Negative & Unsafe & 12 \\
\hline
5 & Pond 2 East Main Drive & Positive & Negative & Unsafe & 12 \\
\hline
6 & Pond 3 East Main Drive & Positive & Negative & Unsafe & 12 \\
\hline
7 & Pond 4 East Main Drive & Negative & Negative & Safe & 12 \\
\hline
8 & Pond 5 East Main Drive & Positive & Negative & Unsafe & 12 \\
\hline
9 & Alpine Spring Water, Crystal Geyser & Negative & Negative & Safe & 10 \\
\hline
10 & Mountain Spring Water, Arrowhead & Negative & Negative & Safe & 10 \\
\hline
11 & Purified Drinking Water, Good and Father & Negative & Negative & Safe & 10 \\
\hline
12 & Clear Alaskan Glacier & Negative & Negative & Safe & 12 \\
\hline
13 & Signature Safeway Purified Drinking Water & Negative & Negative & Safe & 12 \\
\hline
14 & Tap water (Home) & Negative & Negative & Safe & 10 \\
\hline
\end{tabular}
}
\end{table}

To enhance robustness and scalability of the dataset, the previously discussed augmentation algorithm was applied. From each original image, 625 permutations of images were generated. This process resulted in the creation of a field-testing dataset comprising 100,000 microscopic images, all entirely unseen during model development.

\section{Deep Learning Model Construction and Results}

Transfer learning was employed to train a deep learning model. Transfer learning is a machine learning technique, utilizing knowledge acquired from one task to enhance performance in another. For image classification, ResNet18 and ResNet50 are two well-known CNN based models, that have been pre-trained on large data sets and used as a baseline to further train with additional images specific to new problem domains. Both ResNet18 and ResNet50 were originally trained on the ImageNet dataset, a dataset of 1000 classes and over 1,000,000 images. ResNet18 uses 18 learning layers, whereas ResNet50 uses 50 learning layers to train the model \cite{he2016}.

For model development, the ResNet18 and ResNet50 base models were modified using dropout layers and a different fully-connected output layer. Dropout is a regularization technique to improve the performance of CNNs. During training, dropout randomly deactivates the outputs of neurons, effectively preventing information from being propagated to subsequent layers. This approach reduces overfitting and enhances the model's ability to generalize, particularly when encountering unseen data \cite{srivastava2014}. For the model with 18 learning layers (CNN18), a dropout layer with a rate of 0.1 was added after the convolutional layers in each of the third and fourth residual blocks. Also, a dropout layer with a rate of 0.35 was added between the global average pooling layer and the final fully connected output layer. For the model with 50 learning layers (CNN50), layers with higher dropout rates of 0.15 and 0.5 were added at analogous locations. The dropout rates selected for the 50-layer CNN were higher than those selected for the 18-layer CNN because of the 50-layer model's higher processing capacity.

\begin{figure}[H]
\centering
\includegraphics[width=\textwidth]{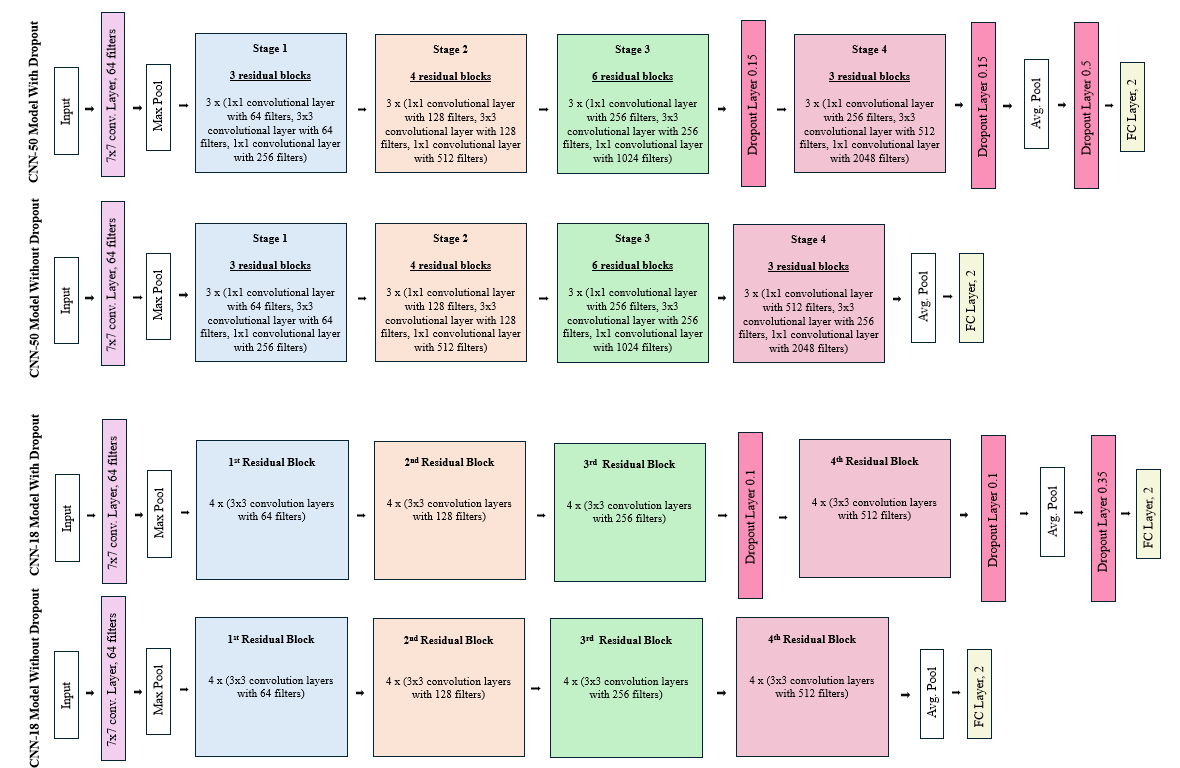}
\caption{Model Architecture of CNN-50 With Dropout, CNN-50 Without Dropout, CNN-18 With Dropout, CNN-18 Without Dropout.}
\label{fig:architecture}
\end{figure}

Prior to training, the foundational data set was split (before augmentation) into three sets. One set, comprising 205 images (approx. 70\%) was the training set. Another set, comprising 61 images (approx. 20\%) was the validation set used during training. Lastly, the remaining 31 images comprised a test set that was not used during training or validation. Then data augmentation was applied on these sets individually to create 19,530 training images, 6,905 validation images, and 3,590 testing images.

The CNN18 and CNN50 models were initialized, and the output layer of the models were each altered to a two class fully connected layer. Thus, instead of 1000 image classes, the models now output 2---safe (negative) and unsafe (positive). Since both CNN18 and CNN50 were originally trained on ImageNet using 224x224-pixel images, the dataset was resized to match the input size requirements -- this was done before feeding the image to the models. The CNN-18 and CNN-50 models were trained and validated using PyTorch \cite{pytorch} on the augmented training and validation sets, both with and without the inclusion of dropout layers. Thus, the models were trained with 19,530 training images and validated on 6,905 images.

\subsection{Model Performance}

When training, a maximum of 50 epochs was used, and the model was terminated when the validation loss did not improve after 5 epochs (i.e., ``patience'' was set to 5 epochs). All models were then tested on the test set constructed from the foundational dataset, as well as on the field-testing dataset. Performance metrics related to the testing are shown in Table \ref{tab:performance_eval} and Table \ref{tab:field_test}. The metrics used are:

\begin{equation}
\text{Accuracy} = \frac{\text{Number of Total Images Correctly Classified}}{\text{Number of Total Images}} \times 100
\end{equation}

\begin{equation}
\text{Precision} = \frac{\text{True Positives}}{\text{True Positives} + \text{False Positives}} \times 100
\end{equation}

\begin{equation}
\text{Recall} = \frac{\text{True Positives}}{\text{True Positives} + \text{False Negatives}} \times 100
\end{equation}

\begin{equation}
\text{F1-score} = \frac{2 \times \text{Precision} \times \text{Recall}}{\text{Precision} + \text{Recall}} \times 100
\end{equation}

\begin{table}[H]
\centering
\caption{Performance during model evaluation.}
\label{tab:performance_eval}
\footnotesize
\adjustbox{max width=\textwidth}{
\begin{tabular}{|l|c|c|c|c|c|c|c|}
\hline
\textbf{Model} & \textbf{Train} & \textbf{Validation} & \textbf{Test} & \textbf{Epochs} & \textbf{Train Acc.} & \textbf{Val. Acc.} & \textbf{Test Acc.} \\
\hline
CNN 50 + Dropout & 19,530 & 6,905 & 3,590 & 3 & 99.97\% & 99.68\% & 99.97\% \\
\hline
CNN 50 & 19,530 & 6,905 & 3,590 & 8 & 100.00\% & 98.48\% & 99.97\% \\
\hline
CNN 18 + Dropout & 19,530 & 6,905 & 3,590 & 1 & 98.41\% & 97.48\% & 76.74\% \\
\hline
CNN 18 & 19,530 & 6,905 & 3,590 & 10 & 100.00\% & 97.55\% & 76.35\% \\
\hline
\end{tabular}
}
\end{table}

\begin{table}[H]
\centering
\caption{Performance in Field Test with 100,000 Unseen Microscopic Images of Unincubated Water.}
\label{tab:field_test}
\footnotesize
\adjustbox{max width=\textwidth}{
\begin{tabular}{|l|c|c|c|c|c|c|c|c|c|}
\hline
\textbf{Model} & \textbf{Acc.} & \textbf{Prec.} & \textbf{Recall} & \textbf{F-1} & \textbf{Opt. Thresh.} & \textbf{Prec. (Opt.)} & \textbf{Recall (Opt.)} & \textbf{Acc. (Opt.)} & \textbf{F-1 (Opt.)} \\
\hline
CNN 50 + Dropout & 92.23\% & 87.37\% & 96.72\% & 91.81\% & 0.8443 & 90.00\% & 94.48\% & 92.79\% & 92.19\% \\
\hline
CNN 50 & 92.49\% & 89.42\% & 94.49\% & 91.88\% & 0.5547 & 90.00\% & 94.22\% & 92.69\% & 92.06\% \\
\hline
CNN 18 + Dropout & 91.80\% & 98.24\% & 85.64\% & 91.51\% & 0.9268 & 90.00\% & 83.89\% & 88.55\% & 86.84\% \\
\hline
CNN 18 & 86.27\% & 81.29\% & 90.26\% & 85.54\% & 0.9844 & 90.00\% & 79.99\% & 86.97\% & 84.70\% \\
\hline
\end{tabular}
}
\end{table}

From Table \ref{tab:performance_eval}, it is seen that the CNN50 models outperform the CNN18 models. This is consistent with expectations since there are more convolution layers in CNN50. Table \ref{tab:field_test} shows the performance on the field test dataset of 100,000 images. The CNN50 models achieve high recall of 94\%-96\%, but precision is slightly lower at around 87\% to 89\%. ``Recall'' refers to the ratio of true positives to the sum of true positives and false negatives, whereas ``Precision'' refers to the ratio of true positives to the sum of true positives and false positives. The lower the recall, the greater the number of false negatives. Similarly, the lower the precision, the greater the number of false positives. When testing water safety, we want false negatives to be as small as possible, as incorrectly classifying unsafe water as safe (i.e., false negatives) can have adverse health consequences. Thus, we prefer a higher recall score than precision. After all, having some false positives only means, the user will take some unnecessary precautions, but not suffer adverse health consequences.

\begin{figure}[H]
\centering
\begin{subfigure}[b]{0.48\textwidth}
\centering
\includegraphics[width=\textwidth]{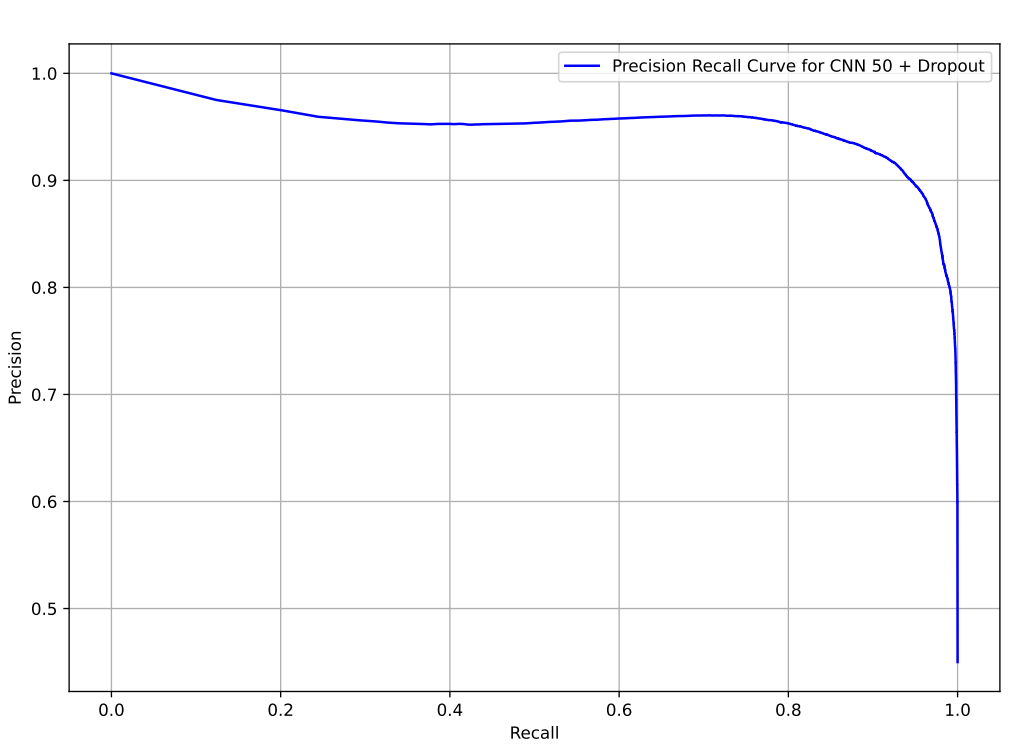}
\caption{CNN50 with dropout}
\end{subfigure}
\hfill
\begin{subfigure}[b]{0.48\textwidth}
\centering
\includegraphics[width=\textwidth]{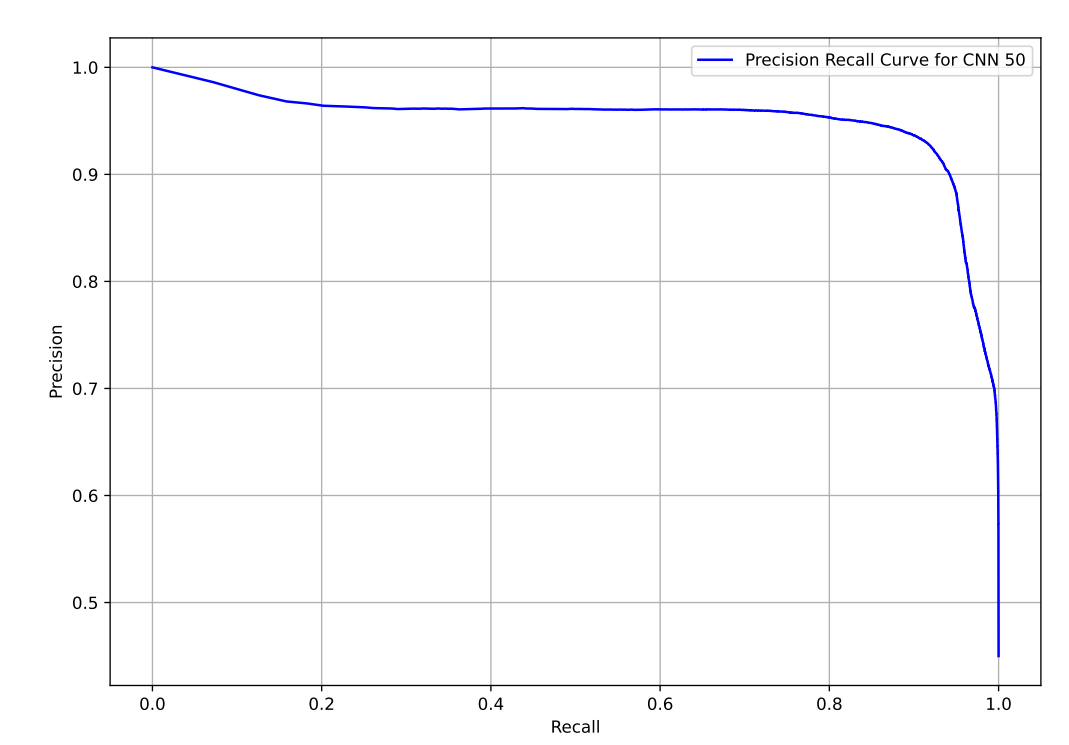}
\caption{CNN50 without dropout}
\end{subfigure}
\caption{Precision Recall Curve for the CNN50 model with dropout (left) and CNN50 model without dropout (right).}
\label{fig:pr_curve_50}
\end{figure}

\begin{figure}[H]
\centering
\begin{subfigure}[b]{0.48\textwidth}
\centering
\includegraphics[width=\textwidth]{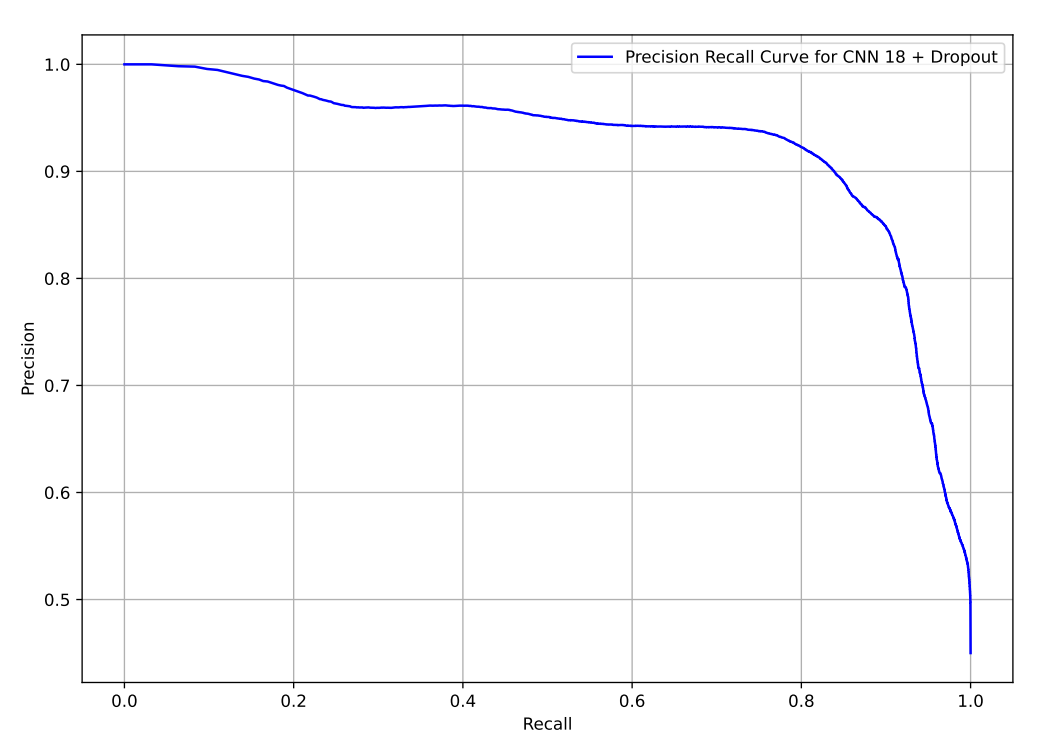}
\caption{CNN18 with dropout}
\end{subfigure}
\hfill
\begin{subfigure}[b]{0.48\textwidth}
\centering
\includegraphics[width=\textwidth]{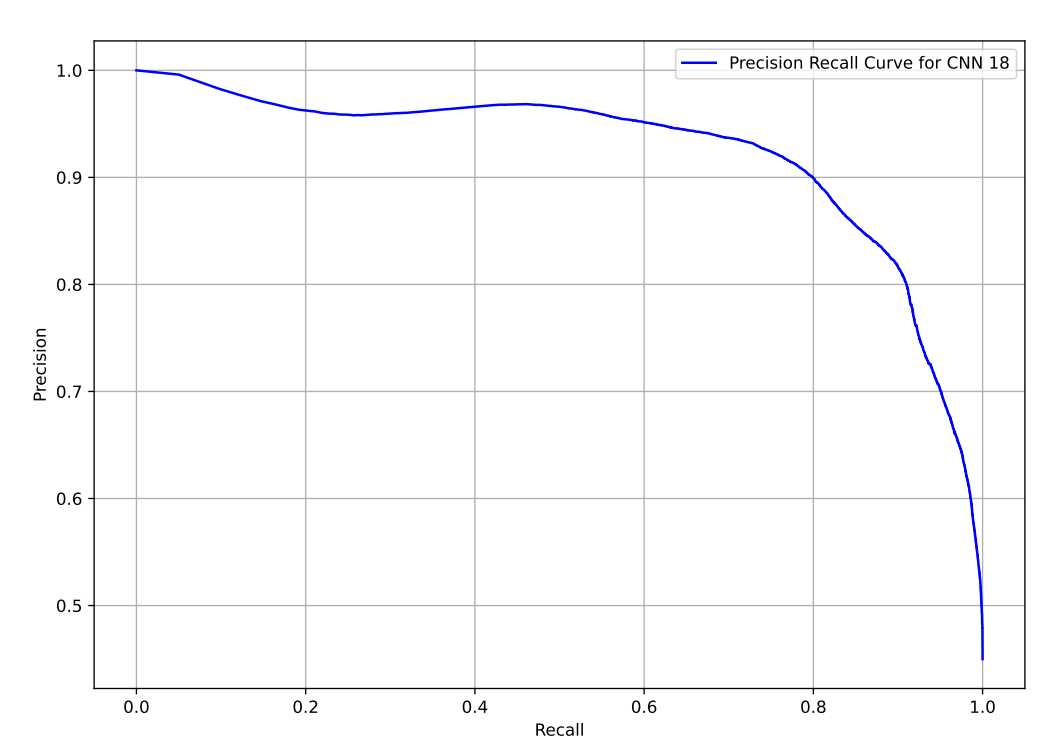}
\caption{CNN18 without dropout}
\end{subfigure}
\caption{Precision Recall Curve for the CNN18 model with dropout (left) and for the CNN18 model without dropout (right).}
\label{fig:pr_curve_18}
\end{figure}

To study the relationship between precision and recall for each of the models, a precision-recall curve was plotted as shown in Figure \ref{fig:pr_curve_50} and Figure \ref{fig:pr_curve_18}. As precision increases, recall generally decreases and vice versa. The different precision-recall combinations were obtained by converting (using a SoftMax layer) the logits output by the deep learning models into probabilities. If the probability exceeded the default threshold probability is 0.5, the sample is classified as unsafe. By changing the threshold value, it is possible to increase or decrease precision and recall. A baseline precision of 90\% was established, and the threshold probability required to achieve this precision was identified. The corresponding recall of the models was then determined based on this threshold. Table \ref{tab:field_test} shows that increasing the precision to 90\% results in a decrease in recall to around 94\% for the CNN50 models. For the CNN18 models, the recall achieved is poor, hence the 18-layer model is not suitable for water safety detection.

The 95\% confidence interval and standard deviation of the different metrics of the models using the optimal thresholds is shown in Table \ref{tab:confidence}. The confidence interval was determined by using a bootstrapping methodology \cite{nakatsu2023}. From the field-testing dataset comprising 100,000 images, resampling with replacement was performed to randomly select 100,000 images. A total of 1,000 such tests were conducted. In each test, the model results for the selected images were sorted, and the lowest and highest 2.5\% of the data points (25 images from each end) were excluded, yielding a 95\% confidence interval.

Overall, the CNN50 model with dropout (the DeepScope model) achieves the highest recall (94.48\% $\pm$ 0.11\%), accuracy (92.79\% $\pm$ 0.08\%), and F-1 Score (92.19\% $\pm$ 0.09\%) with precision set to at least 90\%.

\begin{table}[H]
\centering
\caption{95\% Confidence Intervals and Standard Deviations of Precision, Recall, Accuracy, and F-1 Score of the Models with Optimal Thresholds on Field-testing Dataset.}
\label{tab:confidence}
\footnotesize
\adjustbox{max width=\textwidth}{
\begin{tabular}{|l|c|c|c|c|}
\hline
\textbf{Model} & \textbf{Precision} & \textbf{Recall} & \textbf{Accuracy} & \textbf{F-1 Score} \\
\hline
CNN 50 + Dropout & [0.8973, 0.9027] 0.0014 & [0.9427, 0.9469] 0.0011 & [0.9263, 0.9295] 0.0008 & [0.9199, 0.9236] 0.0009 \\
\hline
CNN 50 & [0.8973, 0.9027] 0.0014 & [0.9402, 0.9443] 0.0011 & [0.9252, 0.9285] 0.0009 & [0.9187, 0.9224] 0.0008 \\
\hline
CNN 18 + Dropout & [0.8972, 0.9029] 0.0014 & [0.8354, 0.8423] 0.0017 & [0.8836, 0.8875] 0.001 & [0.8660, 0.8707] 0.0012 \\
\hline
CNN 18 & [0.8971, 0.9028] 0.0016 & [0.7956, 0.8028] 0.0019 & [0.8676, 0.8718] 0.0013 & [0.8439, 0.8490] 0.0011 \\
\hline
\end{tabular}
}
\end{table}

\section{DeepScope Model Distribution and Results}

\subsection{Model Distribution via Internet}

Integrating a deep learning model into a smartphone application offers significant benefits in terms of accessibility, given that billions of people worldwide have access to smartphones. Furthermore, the model can be periodically updated, with the latest model accessible over the Internet, enabling continuous improvements in performance and accuracy without requiring users to update the application. The feasibility of this model access method was validated through the implementation of a Flask-based web server and an Android-based application that communicates over the Internet \cite{flask, android}.

\begin{figure}[H]
\centering
\includegraphics[width=0.5\textwidth]{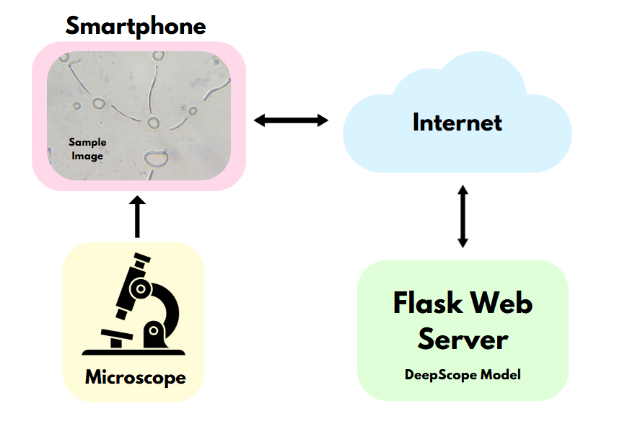}
\caption{Example workflow to test water safety over the Internet using a web server hosting the DeepScope model.}
\label{fig:workflow_internet}
\end{figure}

The web server is hosted on a Linux virtual machine on DigitalOcean, where the necessary packages for running a Flask web server were installed \cite{herbert2021}. The DeepScope model, trained locally and stored as a ``.pth'' file, was uploaded to the remote virtual machine using secure FTP. The Flask server loads the DeepScope model from the ``.pth'' file and remains idle, awaiting incoming requests. Once the Flask server is active on the remote virtual machine, it can be accessed via the Internet through a specific URL. On the Android application, a user interface was developed with a ``Select Image'' button for choosing an image file. After selecting the file, users can click a ``Predict'' button, which sends the image file to the remote server via HTTP POST \cite{herbert2021}.

\begin{figure}[H]
\centering
\includegraphics[width=0.9\textwidth]{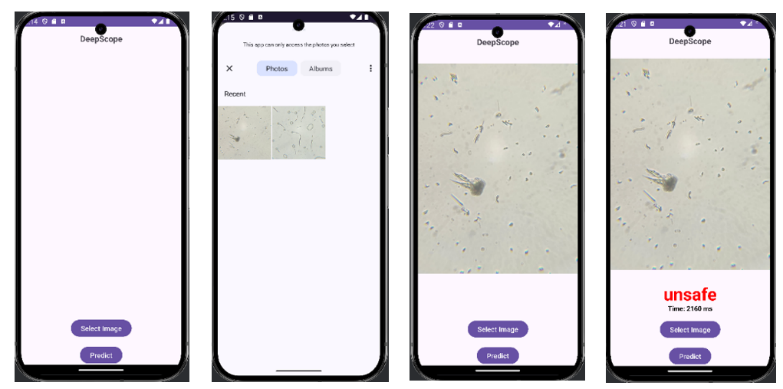}
\caption{Screenshots of DeepScope app executing on an Android device. The app communicates with a Flask server at http://A.B.C.D:5000.}
\label{fig:android_app}
\end{figure}

When the Flask server receives the POST request with the image, the Flask server feeds the image as input after preprocessing. Since the model was trained on square, 224x224-pixel images, the image received over the Internet is cropped and resized to match the input requirements. Subsequently, the DeepScope model classifies the image as ``safe'' or ``unsafe.'' The Flask server returns the prediction to the Android app as a JSON response. When the application was tested, the classification of an image from its selection to the output result was completed in seconds. The workflow of the DeepScope app is depicted in Figure \ref{fig:workflow_internet}, and screenshots of the app user interface are shown in Figure \ref{fig:android_app}. Using such apps, users can assess water quality within seconds of snapping an image of a water sample under a microscope using their smartphone instead of waiting 24-72 hours for results using traditional methods.

This is a greater than 98\% reduction in time compared to the time taken from conventional testing methods such as color-change coliform tests, membrane filtration, etc. (assuming up to 20 minutes time for water sample collection, microscope analysis, and use of the app, compared to an average time of 24 hours for other testing methods).

\subsection{Model Distribution via Smartphone App}

\begin{figure}[H]
\centering
\begin{subfigure}[b]{0.55\textwidth}
\centering
\includegraphics[width=\textwidth]{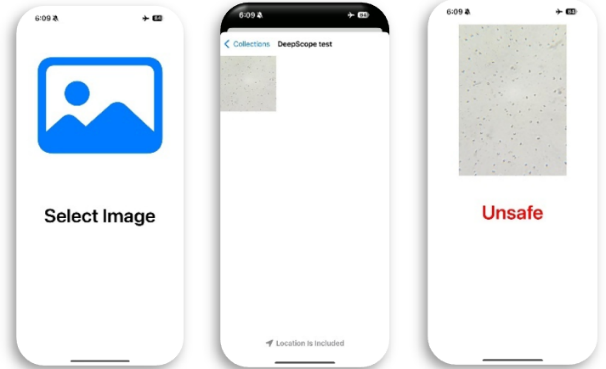}
\caption{iOS app screenshots}
\end{subfigure}
\hfill
\begin{subfigure}[b]{0.4\textwidth}
\centering
\includegraphics[width=\textwidth]{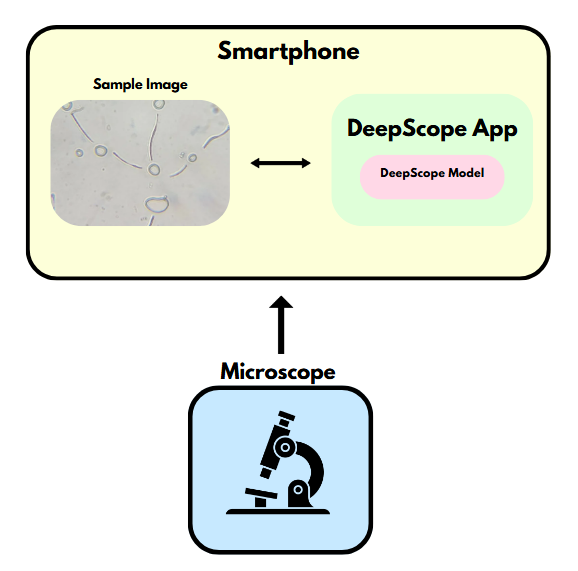}
\caption{Workflow diagram}
\end{subfigure}
\caption{Screenshots and workflow of DeepScope app on iOS device. DeepScope model stored and accessed directly on the smartphone device.}
\label{fig:ios_app}
\end{figure}

While smartphones can access various websites, in some places, Internet connectivity can be a problem. In such cases, it may be preferrable to store the deep learning model on the smartphone itself. To validate this mode of use, a smartphone app was developed where the deep learning model to test water safety is stored and accessed directly on the local device itself. Figure \ref{fig:ios_app} illustrates the workflow when the DeepScope model is stored on the smartphone device. The user snaps an image of the water sample under a microscope using the smartphone and uses a smartphone app to feed the image as input to the DeepScope model stored on the device. The model prediction can then be displayed to the user using the smartphone display.

The application was implemented for iOS devices to validate the distribution of the DeepScope model across both Android and iOS platforms. The PyTorch model was converted into a Core ML ``.mlmodel'' format to enable deployment on iOS devices, which do not support native PyTorch execution. To facilitate this, the model was scripted using torch.jit.script to convert the model into a CoreML format \cite{coreml}. The iOS application was developed using Swift \cite{swift}. To input an image, users interact with a ``Select Image'' button within the interface. The selected image is cropped and resized to 224x224 pixels before being provided as input to the DeepScope model, which classifies the image as either safe or unsafe. The user interface of the application is presented in Figure \ref{fig:ios_app}. Similar to the Android-based application, the iOS app classifies images in seconds.

\section{Discussion and Conclusions}

An innovative system combining microscopy and deep learning to detect water safety was developed. Incubation is not required for this system, as water samples can be directly studied under a microscope, and images captured using smartphone cameras. These images are processed using the DeepScope convolutional neural network model, which is accessible over the Internet or directly on a smartphone device, providing results from microscope images within seconds.

The DeepScope model was developed by training convolutional neural networks using microscope images of bacteria and water samples. A new tile-swap augmentation algorithm was developed to generate thirty-thousand images for model training and testing purposes. Based on the experiments presented, it was concluded that larger numbers of convolution layers in the model improve model accuracy. Additionally, the incorporation of dropout layers in the CNN models was shown to enhance the performance of the inference model.

The DeepScope model was field-tested using microscopic analysis of water collected from 14 different water sources across Sammamish, WA. One hundred thousand images were created from these samples. With 95\% confidence, the DeepScope model achieved a precision of 90.00\% $\pm$ 0.14\%, recall of 94.48\% $\pm$ 0.11\%, accuracy of 92.79\% $\pm$ 0.08\%, and F-1 Score of 92.19\% $\pm$ 0.09\% on the field-test images.

DeepScope greatly exceeds UNICEF's TPP requirements for presence/absence tests, having a 5.52\% false negative rate and 10\% false positive rate. DeepScope detects bacterial contamination in water with a 98\% reduction in time compared to conventional testing methods (color-change coliform tests, membrane filtration tests, etc.) by eliminating the need for pathogen incubation. The safety results were corroborated through parallel color-change coliform testing. These findings suggest that the DeepScope model can reliably detect bacterial contamination in water, demonstrating a high degree of accuracy for field use. Furthermore, Android and iOS apps that can store the DeepScope model or access the model over the Internet were developed. Thus, the DeepScope model can be distributed widely across the globe. The per-test cost of DeepScope is estimated to be \$0.44, since 100 microscope slides and cover slips together cost around \$7, and assuming that a \$90 1000x compound microscope such as AmScope M150 can be used at least 300 times. The cost of using the apps is negligible. This per-test cost is substantially lower than the ideal UNICEF's TPP per-test cost of \$6 \cite{unicef_tpp}. Together, the results indicate that DeepScope is a faster, cost-effective, and highly accurate model that enhances current best-practice water safety testing methods.

\begin{table}[H]
\centering
\caption{Comparison of metrics of different water safety testing methods against ideal UNICEF Target Product Profile}
\label{tab:comparison}
\footnotesize
\adjustbox{max width=\textwidth}{
\begin{tabular}{|l|l|l|l|l|}
\hline
\textbf{Method} & \textbf{DeepScope} & \textbf{Color-Change Coliform} & \textbf{Membrane Filtration} & \textbf{UNICEF TPP Ideal \cite{unicef_tpp}} \\
\hline
\textbf{Performance} & \begin{tabular}[t]{@{}l@{}}5.31-5.71\% FN,\\ 9.73\%-10.27\% FP\\ (95\% CI)\end{tabular} & \begin{tabular}[t]{@{}l@{}}0-23.3\% Error\\ \cite{olstadt2007}\end{tabular} & \begin{tabular}[t]{@{}l@{}}4.3\% FN, 4.3\% FP\\ \cite{astm_ecoli, epa_method1604}\end{tabular} & \begin{tabular}[t]{@{}l@{}}10\% FN, 10\% FP\\ (90\% CI)\end{tabular} \\
\hline
\textbf{Cost/Test} & \$0.44 & \$20 & \$50 & $<$ \$6 \\
\hline
\textbf{Time} & 20 minutes & 24-72 hours & 18-24 hours & $<$ 30 minutes \\
\hline
\end{tabular}
}
\end{table}

Additionally, regarding limitations, DeepScope requires a compound microscope, which may be expensive or inaccessible in limited resource areas. However, cheap compound microscopes with 1000x magnification are no more expensive than the cost of two water safety lab tests. Additionally, since no incubation is performed, there is a risk that no bacteria is visible in the sample being studied, leading to potentially incorrect results. However, the field-testing results indicate this risk is very low as sufficient particles were observed under the microscope in the unsafe samples analyzed.

In cases where very inexpensive and low-quality smartphone cameras are used, it is possible that images may lack sufficient resolution or appear excessively grainy or smudged, leading to inaccurate results. This concern has been mitigated to some extent through the use of dropout during training. Here, during convolution, some portions of the input data are intentionally deactivated, making the model more robust and likely to perform well with lower-quality images. A follow-up study involving a broader range of camera and microscope quality is necessary to thoroughly validate the robustness of the DeepScope models.

As a next step, the DeepScope apps will be made available on smartphone app stores, enabling wider adoption. Additional field testing is intended to be conducted across a broader range of water sources, microscopes, and smartphone cameras. Since covering such diverse geographies, microscope types, and smartphone devices is both costly and time-consuming, a crowdsourcing framework can be developed to obtain microscope images from various locations, microscopes, and smartphones. Furthermore, USEPA or other government approval may be required for the product's public release.

\section*{Acknowledgements and Declarations}

The author would like to thank Dr. Srinivasan Jagannathan for his help in collecting water samples for the field testing described in this paper.

The author does not have any competing interests or conflicts with respect to the contents of this paper. The author received no specific funding for this work. The author confirms sole responsibility for all aspects of the manuscript, including conception, design, data collection, analysis, interpretation, and writing.

\end{document}